\documentclass[amsmath,amssymb,showkeys,showpacs,twocolumn,nobibnotes,aps,prd]{revtex4-1}
\usepackage{graphicx,amsmath,amssymb}
\usepackage{dcolumn}
\usepackage{bm}

\begin{document}

\title{A Self-Consistent Model for Positronium Formation from Helium
Atoms}

\author{E. Ghanbari-Adivi}\email{ghanbari@phys.ui.ac.ir}
\affiliation{Department of Physics and Isfahan Quantum Optics
Group(IQOG), Faculty of Sciences, University of Isfahan, Isfahan
81746-73441, Iran}

\begin{abstract}
  The differential and total cross sections for electron capture by
  positrons from helium atoms are calculated using a first-order
  distorted wave theory satisfying the Coulomb boundary conditions.
  In this formalism a parametric potential is used to describe the
  electron screening in a consistent and realistic manner. The present
  procedure is self consistent because (i) it satisfies the correct
  boundary conditions and post-prior symmetry, and (ii) the potential
  and the electron binding energies appearing in the transition
  amplitude are consistent with the wave functions describing the
  collision system. The results are compared with the other theories
  and with the available experimental measurements. At the considered
  range of collision energies, the results agree reasonably well
  with recent experiments and theories.
\end{abstract}



\maketitle

\section{Introduction}\label{sec:3}
Both positrons ($e^+$) and positronium atoms $(Ps)$ have important
applications in many different branches of physics, chemistry and
other fields~\cite{Charlton1,Surko,Jean,Puska,David,Jerusalem}. This
has motivated numerous studies of collisions in which electrons in
atomic or molecular targets are captured by
positrons~\cite{Gribakin,Fojon,Laricchia,Cooke,Dunlop,Macri1,Macri2,Caradonna,Surdutovich1,Surdutovich2,Sen,Utamuratov,Cheng1,Cheng2,Marler,Le,Nan,Murtagh1,Ke,Kauppila}.
In a collision, each of the various modes into which the system
under study may be fragmented is called a channel. For definiteness,
consider the positron-helium scattering system, on which the present
paper is focused. In the entrance channel of the $e^{+}+He$ system
the positron impacts on the helium atom. The exit channel can be one
of a number of possible fragmentations, such as elastic or
excitation scattering ($e^++He$ and/or $e^++He^*$), positronium
formation ($Ps+He^+$), or ionization ($e^++e^-+He^*$). An exit
channel is said to be open if the corresponding collision is allowed
by all known conservation laws, such as energy and momentum
conservation; otherwise, it is said to be closed. Since the
transition amplitude in general is a function of the energy, the
occurrence probability for each open channel depends on the impact
energy of the projectile. The transition probabilities per unit
time, per unit target scatterer and per unit of the flux of incident
particles with respect to the target are called the cross section.
Experiments measure cross sections, while theoretical studies
usually attempt to compute them. In positron-helium collisions,
since the charge transfer process amounts to a four-body problem,
the theoretical investigation of the rearrangement involves the full
complexities of the quantum mechanical four-body problem and as such
is too difficult to be exactly solved.\par Notwithstanding the
difficulty, numerous theoretical investigations are found in the
literature~\cite{Sen,Utamuratov,Cheng1,Hewitt,Sarkar,Schultz,Mandal,VanReeth,Fojon2,Igarashi,Campbell},
along with experimental
studies~\cite{Caradonna,Charlton2,Diana,Fromme,Fornari,Overton,Murtagh2}.
Most of the theories were formulated within a single active electron
picture. In this formalism, it is important to provide a consistent
and realistic treatment of the passive electron screening effects,
in order to satisfy the Coulomb boundary conditions and post-prior
equivalency and to make the wave functions, binding energies, and
Coulomb phase factors consistent with each other.\par In the context
of K-shell electron capture processes by fast protons from multi
electron atoms, Decker and Eichler have discussed the deviations of
a few of the standard formalisms from these constraints in some
detail~\cite{Decker}. They adopted a parameterized potential due to
Green, Sellin and Zocher~\cite{Green1,Green2}, the GSZ potential, to
construct a self-consistent screened boundary-corrected
first-Born-approximation theory for calculation of the K-shell
electron capture total cross sections in collisions of proton with
helium and carbon as well as collision of alpha particles with
lithium atoms.\par In the present paper a distorted-wave
boundary-corrected first-order Born (DWB1B) formalism accompanied by
the GSZ potential is applied to obtain a satisfactory description of
positronium formation in positron-helium atom collisions at
intermediate energies. In this model, the single-zeta Hartree-Fock
wave function and its corresponding binding energy, which show very
good agreement with those for GSZ potential, describe the initial
bound state of the electron. Consequently the wave function and the
corresponding energy in the post or prior formalisms are the
eigenfunction and eigenenergy of the same effective screening
potential that appears explicitly in the corresponding amplitudes.
Thus, in addition to satisfying both the Coulomb boundary conditions
an post-prior symmetry, the wave functions, binding energies and
Coulomb phase factors are consistent with each other.\par The plan
of the paper is as follows. Section~2 outlines the formalism.
Section 3 presents and discusses the results and compares them with
other theories and experimental data. The concluding remarks and a
summary comprise the last section. Unless otherwise stated, atomic
units are used throughout.

\section{Theory}\label{sec:2}
Consider a bare ion $P$ with mass $M_P$ and charge $Z_P$ impingent
on an atomic target composed of an active electron $e$ and a
residual target ion $T$ with masses $m$ and $M_T$, respectively.
During the collision, the projectile captures the electron.
According to the time-independent boundary-corrected perturbation
formalism of electron capture developed by Toshima et
al~\cite{Toshima} and Belki\'{c} et al~\cite{Belkic}, the
first-order corrected-boundary distorted-wave Born(DWB1B) amplitude
for such a process has the prior and post forms
\begin{equation}\label{Eq1}
  {\cal A}_{DWB1B}^{(prior)}= \left\langle\chi_f\left|(V_i-V_i^\infty)\right|\chi_i\right\rangle
\end{equation}
and
\begin{equation}\label{Eq2}
  {\cal A}_{DWB1B}^{(post)}=
  \left\langle\chi_f\left|(V_f-V_f^\infty)\right|\chi_i\right\rangle,
\end{equation}
where $V_i^\infty$ and $V_f^\infty$ are the asymptotic limits of the
distorting potentials $V_i$ and $V_f$ in initial and final channels.
In position space, the corresponding distorted wavefunctions
$\chi_i$ and $\chi_f$ are given by the equalities
\begin{equation}\label{Eq3}\begin{split}
\chi_i&({\bf r}_T,{\bf R}_T,{\bf R})=\phi_i({\bf r}_T)\exp(i{\bf
K_i}\cdot{\bf R}_T)\\
&\qquad\times \exp\left(i{Z_P(Z_T^a-1)\over {\rm v}_i}\ln({\rm
v}_iR-{\bf v}_i\cdot{\bf R})\right)\end{split}
\end{equation}
and
\begin{equation}\label{Eq4}\begin{split}
\chi_f&({\bf r}_P,{\bf R}_P,{\bf R})=\phi_f({\bf r}_P)\exp(i{\bf
K_f}\cdot{\bf R}_P)\\ &\qquad\times\exp\left(i{Z_T^a(Z_P-1)\over
{\rm v}_f}\ln({\rm v}_f R-{\bf v}_f\cdot{\bf R})\right),\end{split}
\end{equation}
in which $\phi_i({\bf r}_T)$ and $\phi_f({\bf r}_P)$ are the initial
and final bound-state electronic wave functions. In Eqs.~(\ref{Eq3})
and~(\ref{Eq4}), the plane wave functions describe the
heavy-particle motion in the entrance and exit channels, and the
phase factors ensure that the overall solutions satisfy the proper
asymptotic boundary conditions for the distorting potentials.  In
these equations, ${\bf
  r}_P$ and ${\bf r}_T$ are the electron coordinates relative to the
projectile and target nucleus, respectively, ${\bf R}_P$ is the
position vector of the center of mass of the $Pe$ subsystem relative
to $T$, ${\bf R}_T$ is a similar vector directed from the center of
mass of the $Te$ subsystem to $P$, and $\bf R$ is the internuclear
coordinate vector, directed from $T$ to $P$. The bare projectile and
the asymptotic target-ion charges are $Z_P$ and $Z_T^a$,
respectively, ${\bf v}_i$ and  ${\bf v}_f$ are the projectile
velocity in the initial channel and  resultant bound subsystem
velocity in the final channel, respectivley, and ${\bf K}_i$ and
${\bf K}_f$ are the wave vectors describing the relative motions in
the two channels.\par The Coulomb interaction between the active
electron and the projectile ion is denoted $V_{Pe}({\bf
r}_P)=-Z_P/r_P$, with the asymptotic form of $V_{Pe}^\infty({\bf
R})=-Z_P/R$. We take the internuclear potential to be that of a bare
nucleus screened by the nonactive electrons represented by
\emph{frozen} orbitals. For this purpose, the parameterized
potential obtained by Green, Sellin and Zocher
(GSZ)~\cite{Green1,Green2} is adequate because it has the correct
behavior near the nucleus and at infinity and because it yields
orbital binding energies and wave functions in good agreement with
the corresponding Hartree-Fock results~\cite{Clementi}. Within a
single active electron model, using the GSZ potential for a neutral
atom, one can write the projectile-target interaction in coordinate
representation, $V_{PT}({\bf R})$ as
\begin{equation}\label{Eq5}
V_{PT}(R)={Z_P\over R}\left[1+(Z_T-1)\Omega(R)\right],
\end{equation}
where $Z_T$ is the nuclear charge of the multielectron atomic target
and $\Omega(R)$ is given by the expression
\begin{equation}\label{Eq6}
\Omega(R)={1\over 1-H[1-\exp{(R/d)}]}.
\end{equation}
Optimized values for the characteristic parameters $H$ and $d$ have
been given by Szydlik and Green~\cite{Green2} for $Z_T\le 18$. Near
the nucleus the projectile ion feels the bare nuclear charge $Z_T$,
while at large distances the potential behaves as $Z_P/R$
corresponding to the residual singly charged ion.\par With the
potential~(\ref{Eq5}), the post and prior amplitudes~(\ref{Eq1})
and~(\ref{Eq2}) satisfy the appropriate boundary conditions in the
initial and final channels. It is easy to see that the two
amplitudes are identical provided that $\phi_i({\bf
  r}_T)$ and $\phi_f({\bf r}_P)$ be the exact solutions of the atomic
Schr\"{o}dinger equations associated with the potentials
$V_{Te}({\bf   r}_T)$ and $V_{Pe}({\bf r}_P)$, respectively. The
wave functions, binding energies and Coulomb phase factors are
moreover consistent with each other. Hereafter, given the
equivalence between the prior and post wavefunctions, we will omit
the superscripts \emph{post} and \emph{prior} in Eqs.~(\ref{Eq1})
and~(\ref{Eq2}) and write ${\cal   A}_{DWB1B}$ to denote the
amplitudes.\par The Fourier transform of the $V_{PT}$ potential can
be written in the form
\begin{equation}\label{Eq7}
\tilde{V}_{PT}({\bf k})=\sqrt{2\over \pi}{Z_P\over k^2}+\sqrt{2\over
\pi}{Z_P(Z_T-1)}\tilde{\cal{V}}(k),
\end{equation}
where
\begin{equation}\label{Eq8}
\tilde{\cal{V}}(k)={1\over k}\int_0^\infty dR\Omega(R)\sin(kR).
\end{equation}\par
With the given interaction potentials, $V_{Pe}({\bf r}_P)$ and
$V_{PT}({\bf R})$, the transition amplitude for electron capture
comprises three terms:
\begin{equation}\label{Eq9}
\mathcal{A}_{DWB1B}={\cal A}_1+{\cal A}_2+{\cal A}_3,
\end{equation}
with the following explicit expressions for the partial amplitudes
${\cal A}_1$, ${\cal A}_2$ and ${\cal A}_3$ in coordinate
representation:
\begin{equation}\label{Eq10}\begin{split}
{\cal A}_1=&-Z_P\int d{\bf r}_P d{\bf R}_P \chi^*_f({\bf r}_P,{\bf
R}_P,{\bf R}){1\over r_P}
\chi_i({\bf r}_T,{\bf R}_T,{\bf R})\\
{\cal A}_2=&+Z_P\int d{\bf r}_P {\bf R}_P \chi^*_f({\bf r}_P,{\bf
R}_P,{\bf R}){1\over R}
\chi_i({\bf r}_T,{\bf R}_T,{\bf R})\\
{\cal A}_3=&Z_P(Z_T-1)\int d{\bf r}_P d{\bf R}_P \chi^*_f({\bf
r}_P,{\bf R}_P,{\bf R}){\Omega(R)\over R}\\
&\qquad\qquad\qquad\times \chi_i({\bf r}_T,{\bf R}_T,{\bf
R}).\end{split}
\end{equation}\par
We now consider the special case $Z_P=1$ and the transition from an
initial $1s$ hydrogen-like state such as $\phi_i({\bf x}_T)=N_T
\exp(-\zeta_T x_T)$ to a final $1s$ hydrogen-like state of form
$\phi_f({\bf
  x}_P)=N_P \exp(-\zeta_P x_P)$, where $N_P$ and $N_T$ are
normalization factors.  We apply the Scr\"{o}dinger equation for the
final bound subsystem, use the Fourier transform analysis, and
evaluate the resulting integrals to obtain the following expression
for the first partial amplitude, ${\cal A}_1$:
\begin{equation}\label{Eq11}\begin{split}
{\cal A}_1=&-16\pi^2N_PN_TZ_P({K^2\over
2\mu_f}-\epsilon_f)\\
&\qquad\qquad\times{\partial^2\over
\partial\zeta_P\partial \zeta_T}\left[{1\over
(K^2+\zeta_P^2)(J^2+\zeta_T^2)}\right],\end{split}
\end{equation}
where $\mu_f=M_Pm/(M_P+m)$ is the reduced final-state mass of the
bound subsystem, and ${\bf J}$ and ${\bf K}$ are the momenta
transferred to the target ion and the projectile, respectively,
during the collision.\par Similarly, using the Fourier transform
techniques and Lewis integral~\cite{Lewis}, we derive the following
closed form for the second partial amplitude ${\cal A}_2$:
\begin{equation}\label{Eq12}\begin{split}
&{\cal A}_2=8\pi^2N_PN_TZ_P\\
&\ \times {\partial^2\over
\partial\zeta_P\partial \zeta_T}\left[(\alpha^2-\beta)^{-1/2}\ln
\left[{\alpha+(\alpha^2-\beta)^{1/2}\over\alpha-(\alpha^2-\beta)^{1/2}}\right]\right],\end{split}
\end{equation}
with
$$\alpha=(K^2+\zeta_P^2)\zeta_T+(J^2+\zeta_T^2)\zeta_P$$
and
$$\beta=(K^2+\zeta_P^2)(J^2+\zeta_T)[v_f^2+(\zeta_P^2+\zeta_T^2)].$$\par
Finally, we use Fourier analysis to write the third partial
amplitude ${\cal A}_3$ in the form
\begin{equation}\label{Eq13}
{\cal A}_3=8\pi^2 N_P N_T Z_P(Z_T-1){\partial^2\over
\partial \zeta_P \partial \zeta_T} {\cal I}(\zeta_P,\zeta_T),
\end{equation}
where
\begin{equation}\label{Eq14}
{\cal I}(\zeta_P,\zeta_T)= {1\over\pi^2}\int d{\bf k} {\tilde{\cal
V}(k)\over |({\bf k}-{\bf K}|^2+\zeta_P^2) ({\bf k}+{\bf
J}|^2+\zeta_T^2)}.
\end{equation}\par
Before numerical calculations become possible, we have to
analytically simplify the right-hand side of Eq.~(\ref{Eq13}). To
this end, notice taken that $\tilde{\cal V}(k)$ is a radial
function, we use Feynman's identity~\cite{Feynman} to evaluate the
angular integral over ${\bf k}$:
\begin{equation}\label{Eq15}
{\cal I}(\zeta_P,\zeta_T)=-{1\over\pi}\int_0^1
{dx\over\gamma}\int_0^\infty dR \Omega(R)\int_{-\infty}^{+\infty}dk
{\sin(kR)\over k^2+2\gamma k+\delta},
\end{equation}
where
$$\gamma=\sqrt{J^2(1-x)^2+K^2x^2-2x(1-x){\bf J}\cdot{\bf K}},$$
and
$$\delta=(J^2+\zeta_T^2)(1-x)+(K^2+\zeta_P^2)x.$$\par
Next, we apply Cauchy's residue theorem to the last integral on the
right-hand side of Eq.~(\ref{Eq15}) to see that
\begin{equation}\label{Eq16}
\int_{-\infty}^{+\infty}dk {\sin(kR)\over k^2+2\gamma
k+\delta}=-{\pi\over \sqrt{\delta-\gamma^2}} {\rm
e}^{-R\sqrt{\delta-\gamma^2}}\sin(\gamma R).
\end{equation}\par
With this, we have reduced the four-dimensional integral ${\cal
  I}(\zeta_P,\zeta_T)$ in Eq.~(\ref{Eq14}) to a two dimensional form,
\begin{equation}\label{Eq17}
{\cal I}(\zeta_P,\zeta_T)=\int_0^1
{dx\over\gamma\sqrt{\delta-\gamma^2}}\int_0^\infty dR \Omega(R){\rm
e}^{-R\sqrt{\delta-\gamma^2}}\sin(\gamma R),
\end{equation}
which can be easily computed numerically, along with the derivatives
on the right-hand side of Eq.~(\ref{Eq13}).\par For single-electron
capture in impact of positron on helium atoms $Z_T=2.0$,
$\epsilon_i=-0.89648$, $\epsilon_f=-0.25$, $\mu_f=0.5$,
$\zeta_P=0.5$, $\zeta_T=1.6875$, $N_P=\zeta_P\sqrt{\zeta_P/\pi}$,
$N_T=\zeta_T\sqrt{\zeta_T/\pi}$ and the corresponding differential
and total cross sections are given by
\begin{equation}\label{Eq18}
\sigma(\theta)={d\sigma\over d\Omega}=\frac{2}{\pi
^2}\frac{v_f}{vi}|{\cal A}_{DWB1B}|^2,
\end{equation}
and
\begin{equation}\label{Eq19}
\sigma_{Total}=2\pi\int_0^\pi \sigma(\theta)\sin\theta d\theta,
\end{equation}
respectively.\par To account for the excited final states, we
multiply the right-hand side of Eq.~(\ref{Eq19}) by 1.202, according
to the Oppenheimer $n^{-3}$ scaling rule~\cite{Oppenheimer}.
\begin{figure}
\begin{minipage}{0.5\linewidth}
\includegraphics[scale=0.4]{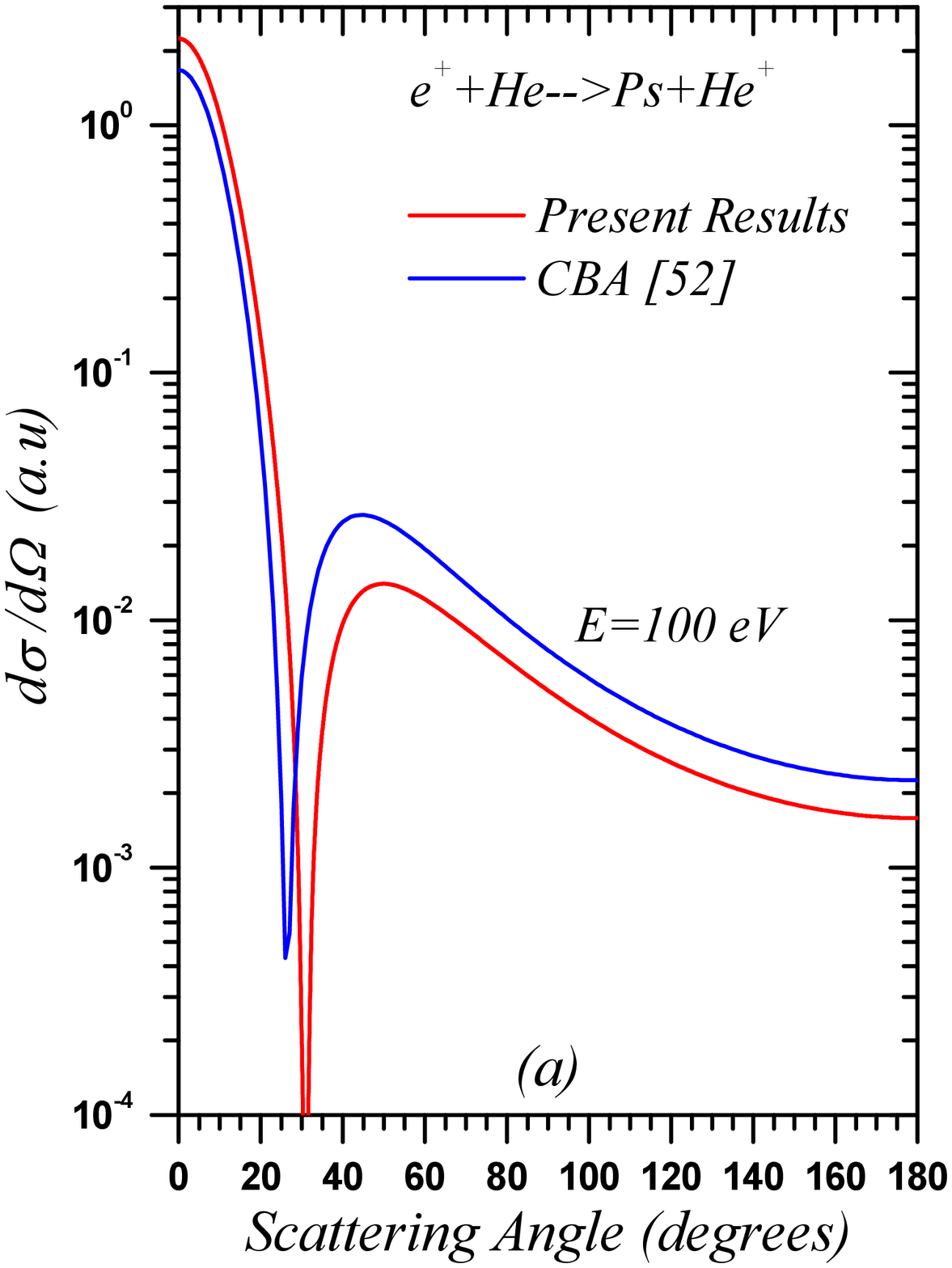}
\end{minipage}
\begin{minipage}{0.5\linewidth}
\includegraphics[scale=0.4]{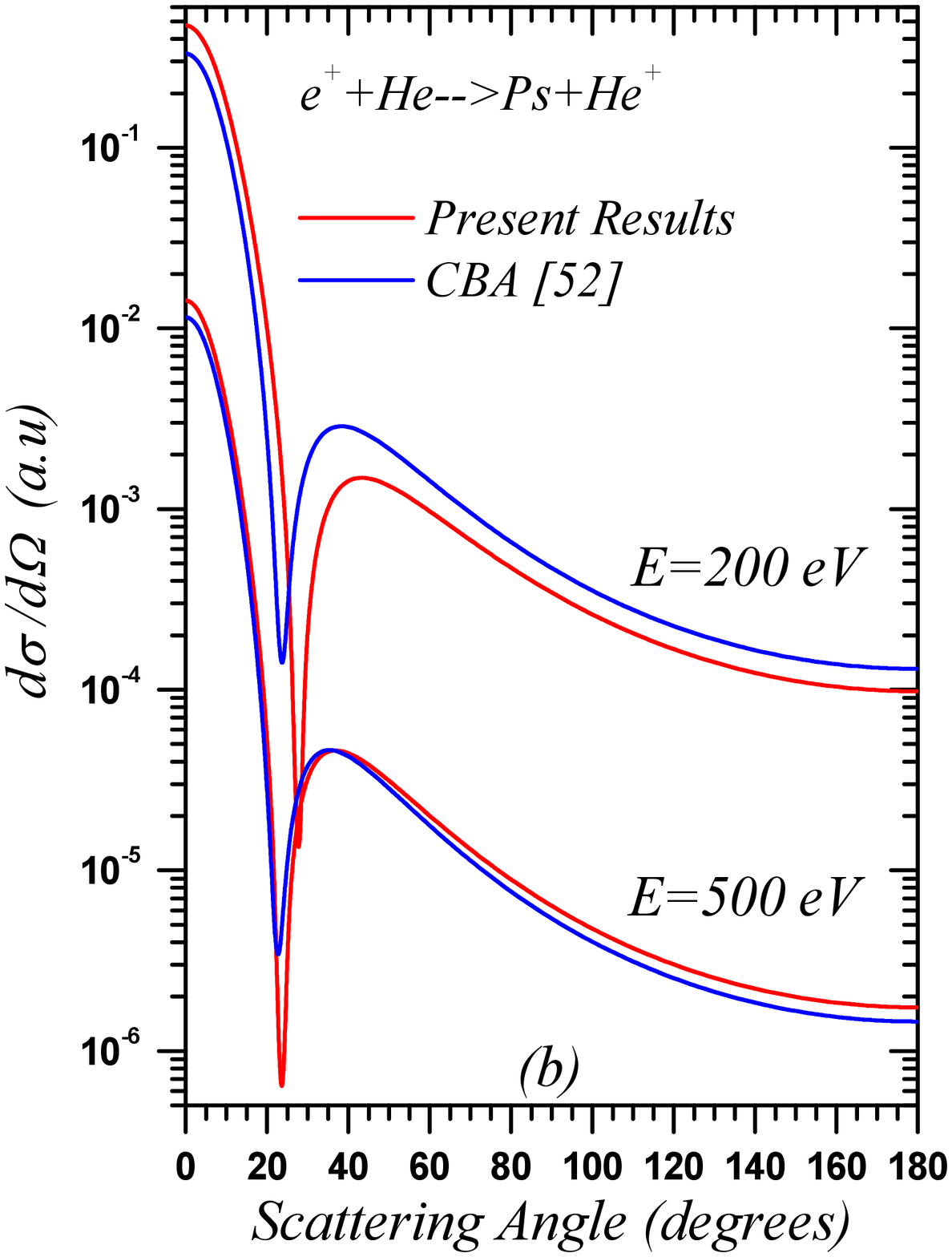}
\end{minipage}
  \caption{Angular distribution of the differential cross sections for
  Ps formation from helium atoms.}\label{Fig1}
\end{figure}
\section{Results and Conclusions}\label{sec:3}
We now present the computed differential and total cross sections
for positronium formation in positron collisions with helium atoms
and compare them with other theoretical approaches and experimental
data. In order to evaluate the third partial transition amplitude,
${\cal A}_3$, we rely on the Gaussian quadrature method to compute
the two-dimensional integral ${\cal I}(\zeta_P,\zeta_T)$ in
Eq.~(\ref{Eq17}).\par Figure~\ref{Fig1} shows the angular
differential cross sections for positronium formation at three
incident energies, 100, 200 and 500~eV. All three curves display a
dip due to the cancelation between the contributions of the
attractive and repulsive interactions to the first-order transition
amplitude. A similar, unphysical dip has been found in a first-order
Born treatment of electron capture in the collision of a proton with
hydrogen atoms~\cite{Band}. Figures~\ref{Fig1}(a) and (b) compare
our results with those of the Coulomb-Born approximation
(CBA)~\cite{Ghanbari} for three impact energies: 100, 200 and
500\,eV. In our computation, the dark angles, i.~e., the angles at
which cancelation occurs, are $31^\circ$, $28^\circ$ and $24^\circ$,
while the CBA displays dark angles of $26^\circ$, $24^\circ $ and
$23^\circ$, respectively. In both treatments, the dark angle becomes
smaller as the incident energy grows.\par Our results differ
significantly from the CBA data at 100~eV incident energy. Below
(above) $30^\circ$ scattering angle our differential cross sections
are always smaller (larger) than the CBA results. As the impact
energy increases, the difference between the our results and the CBA
results shrinks. Compared with the difference at 100~eV, the
difference at 200~eV is smaller, and the difference at 500~eV is
negligible. For energies above 500~eV, our results are consistent
with the CBA results. Consequently, the total cross sections
obtained from the two formalisms converge at high energies.
\begin{figure}
\includegraphics[scale=0.4]{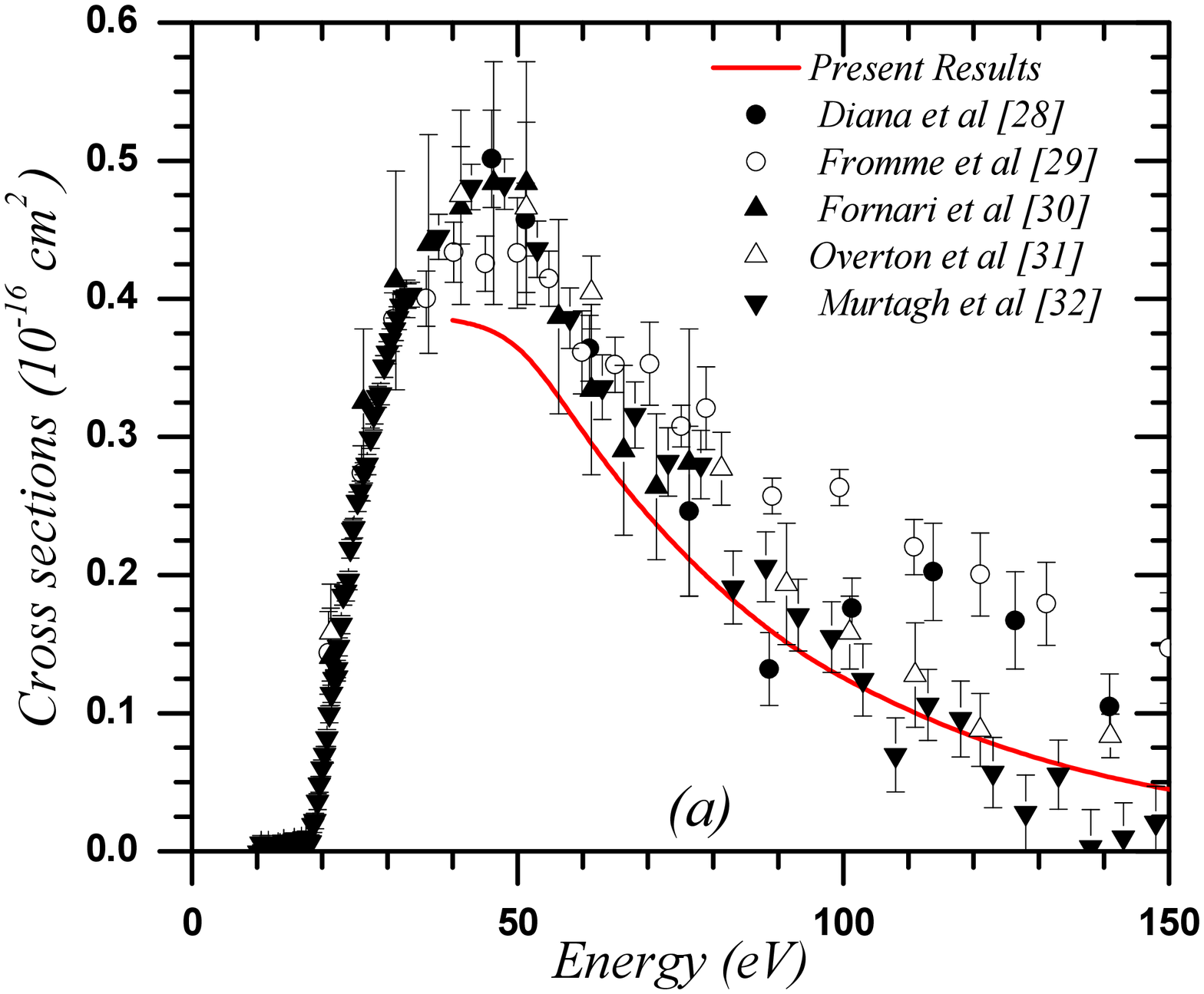}
\includegraphics[scale=0.4]{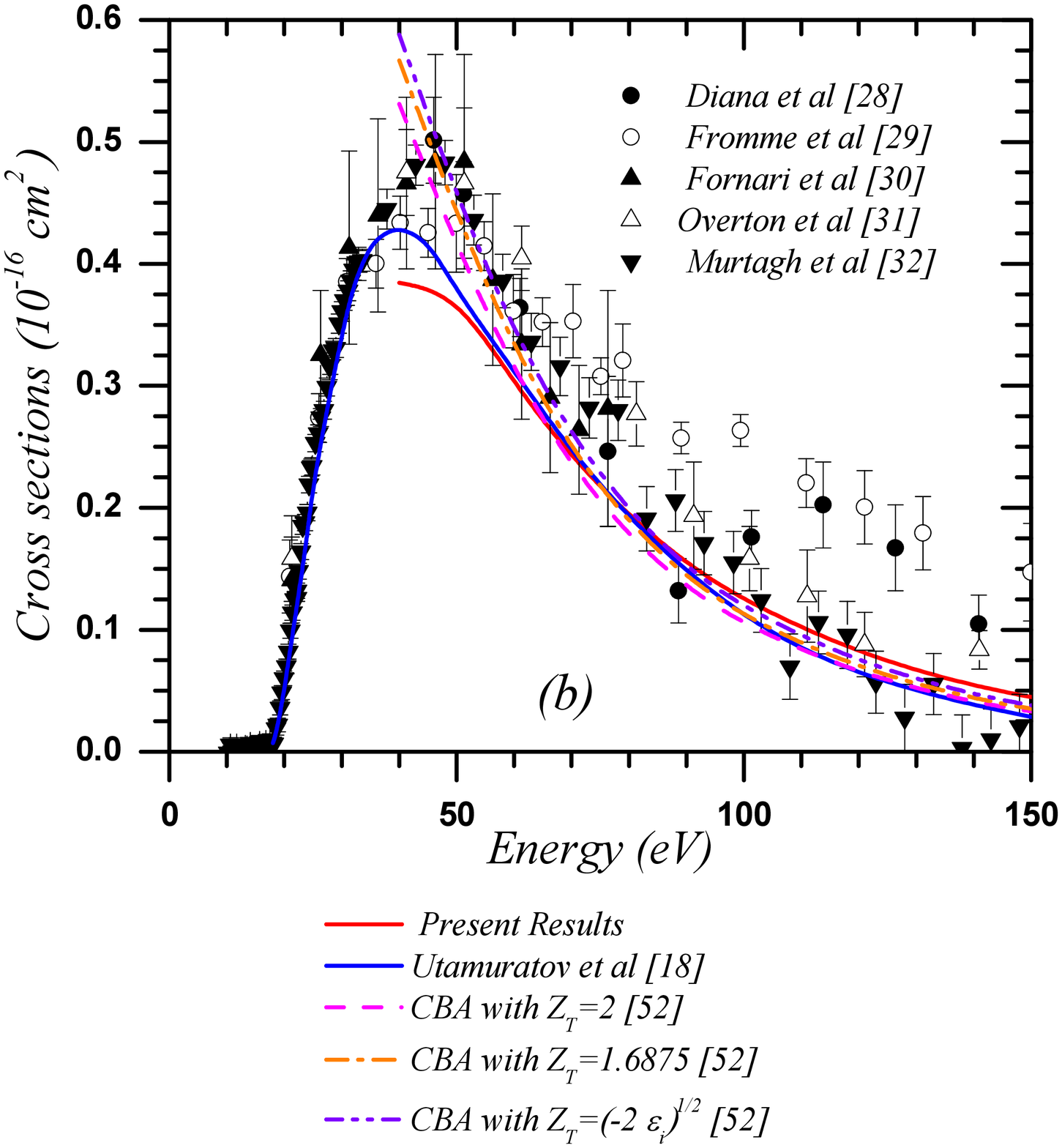}
\caption{Total cross sections for Ps formation in collisions of
  positrons with helium atoms, compared with the results of the listed
  theoretical and experimental studies.}\label{Fig2}
\end{figure}
Figure~\ref{Fig2} compares the total positronium-formation cross
sections obtained from our formalism with the experimental
measurements of Diana~et~al~\cite{Diana},
Fromme~et~al~\cite{Fromme}, Fornari~et~al~\cite{Fornari},
Overton~et~al~\cite{Overton} and Murtagh~et~al~\cite{Murtagh2} and
also with the results of the $CBA$ with $Z_T = 1.6875$, 2 and
$\sqrt{-2\epsilon_i}$~\cite{Ghanbari} and of the two-center
convergent close-coupling (CCC) formalism~\cite{Utamuratov}. As
Fig.~2(a) shows, in spite of its simplicity our formalism yields
reasonable agreement with experiment at energies above 50~eV.  Both
the present approach and the CBA formalism~\cite{Ghanbari} are based
on perturbation expansion, an approach long known to be unreliable
at low scattering energies.\par
  For the first-order Born approximation to be accurate, the
higher-order terms in the expansion series must be small in
comparison with the first-order term. At high energies, since
two-step scattering processes, such as the Thomas double-scattering
mechanism, become dominant at high energies, the first-order
approach yields poor approximations to the differential cross
sections. All considered, we see that first-order methods are only
reliable in a bracket of energies not so low as to render the
perturbative method inapplicable and not so high as to make
double-scattering mechanisms dominant.\par
  Both the CBA and our approach satisfy the correct boundary
conditions. The two methods are nonetheless nonequivalent: while the
CBA breaks post-prior symmetry, our approach is symmetric, so that
the initial and final wavefunctions associated with GSZ potential
are consistent with the scattering potential and bound energies. Our
approach is hence expected to be more appropriate to describe
experimental processes. As the plots in Fig.~2(b) show, however,
only small differences separate the CBA results from ours.\par
  More accurate positronium-formation cross sections are obtained with
the CCC formalism, a more elaborate method employing a
multiconfigurational wavefunction treatment to describe the He
ground state. The accuracy of the calculated cross sections is
controlled by the size of the basis and of the set quantum numbers
of the included states at each center. Figure~\ref{Fig2}(b) shows
good agreement between the CCC curve and the experimental data.\par
  Within the energy bracket in which it is reliable, the solid line
depicting our results in Fig.~\ref{Fig2}(b) agrees with the CCC
line. The difference between the two sets theoretical data is less
than $0.7\%$ around 50~$eV$, while deviations of approximately 17\%
separate the CBA and CCC curves in the same region. At higher
energies the results of the three formalisms converge and display
good agreement with the experimental data.\par
\begin{figure}
\includegraphics[scale=0.4]{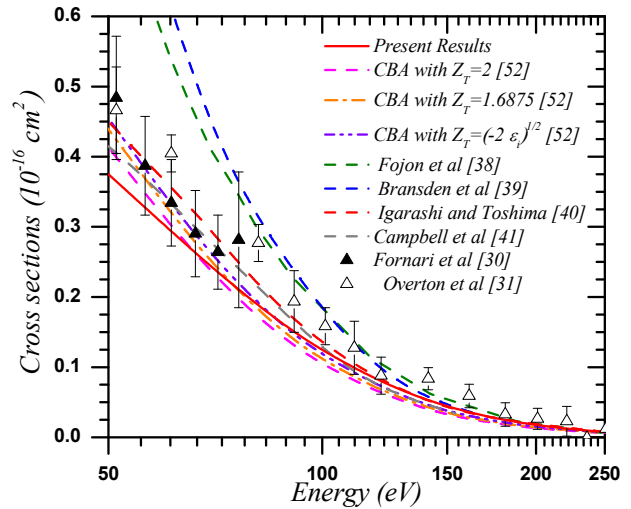}
\caption{Total cross sections, compared with the results of the
CBA~\cite{Ghanbari}, B1B~\cite{Fojon2}, DW~\cite{Bransden},
CDW~\cite{Igarashi} and CC~\cite{Campbell} procedures and with the
experimental data from Fornari~et~al~\cite{Fornari} and
Overton~et~al~\cite{Overton}.}\label{Fig3}
\end{figure}
\begin{figure}
\includegraphics[scale=0.4]{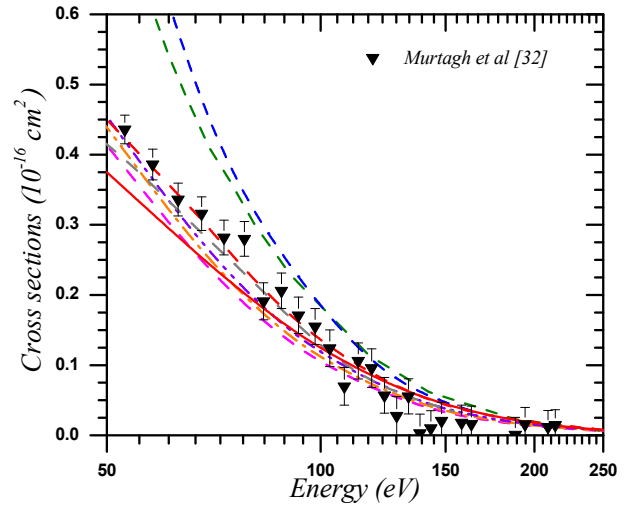}
\caption{Theoretical cross sections, as in Fig.~\ref{Fig3}, compared
  with the measurements by Murtagh et al~\cite{Murtagh2}}\label{Fig4}
\end{figure}
Figure~\ref{Fig3} compares various theoretical approaches with
experiment. Cross sections calculated by the CBA with $Z_T=1.6875,\
2$ and $\sqrt{-2\epsilon_i}$~\cite{Ghanbari}, the correct-boundary
first-order Born~(B1B)~\cite{Fojon2} formalism, the distorted-wave
Born~(DW) approximation~\cite{Bransden}, the continuum distorted
wave~(CDW) approximation~\cite{Igarashi} and the close-coupling
method~(CC)~\cite{Campbell} are compared with the experiments of
Fornari~et~al~\cite{Fornari} and Overton~et~al~\cite{Overton}. The
CBA, CDW, and CC agree well with the measurments in the depicted
energy range. The agreement between B1B and DW at high energies is
fair, but at lower energies those theories deviate from the
measurements.\par Our predictions also deviate from experiment at
low energies, but the agreement improves as the energy increases.
Compared with the results of B1B and DW, our results are closer to
the experimental data at low energies.\par Figure~\ref{Fig4}
compares the above discussed theories with the earlier experimental
measurements reported by Murtagh~et~al~\cite{Murtagh2}. The CDW, CC,
and CBA formalisms yield results in very good agreement with
experiment over the entire energy range. Our results are also in
good agreement at energies above 75~eV, but the discrepancies are
sizable at lower energies. The B1B and DW results are above the data
by Murtagh~et~al~\cite{Murtagh2} for all energies lower than 200~eV,
the deviation growing considerably as the impact energy decreases.
Above 200~eV, all theories converge to the experimental data.\par In
addition to the CCC formalism, a number of theoretical descriptions
of positronium formation in positron-helium atom collisions have
been recently reportde, on the basis of the momentum-space
coupled-channel optical (CCO) method \cite{Cheng1} and second-order
distorted-wave approximation (DWA) \cite{Sen}. Figure 5 compares our
results in the energy range of 50-250~eV with the available results
from the CCO and DWA formalisms, as well as with the experimental
data reported by different
groups~\cite{Diana,Fromme,Fornari,Overton,Murtagh2}. Our first-order
distorted wave results are very close to the predictions of the
second-order DWA, in which the cross sections for Ps formation in
the 1s orbital has been added to 1.66 times the Ps cross sections in
the $n=2$ orbitals to yield the total Ps formation cross
sections.\par Below 80 eV our results understimate the experimental
cross sections. At small energies all theories in the plot become
inaccurate, the the CCO cross sections overshooting the data by
Murtagh et al~\cite{Murtagh2}. At relatively higher energies, above
80~eV our results agree with the other theoretical predictions and
with the measurements.\par
\begin{figure}
\includegraphics[scale=0.4]{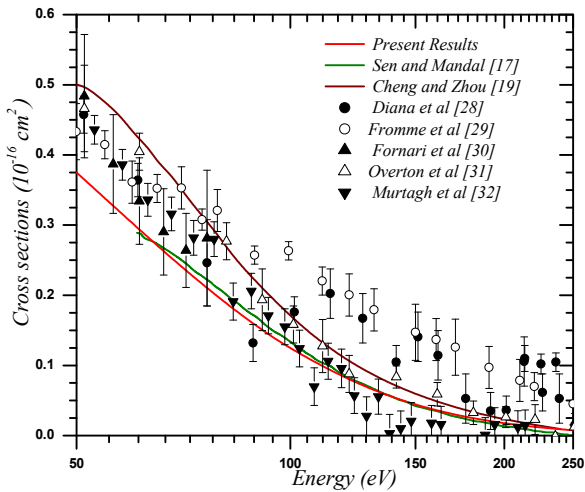}
\caption{Cross sections for positronium formation in helium atoms
compared with all available experimental data and with the results
of the DWA~\cite{Sen} and CCO~\cite{Cheng1} formalisms.}\label{Fig5}
\end{figure}
\begin{figure}
\includegraphics[scale=0.4]{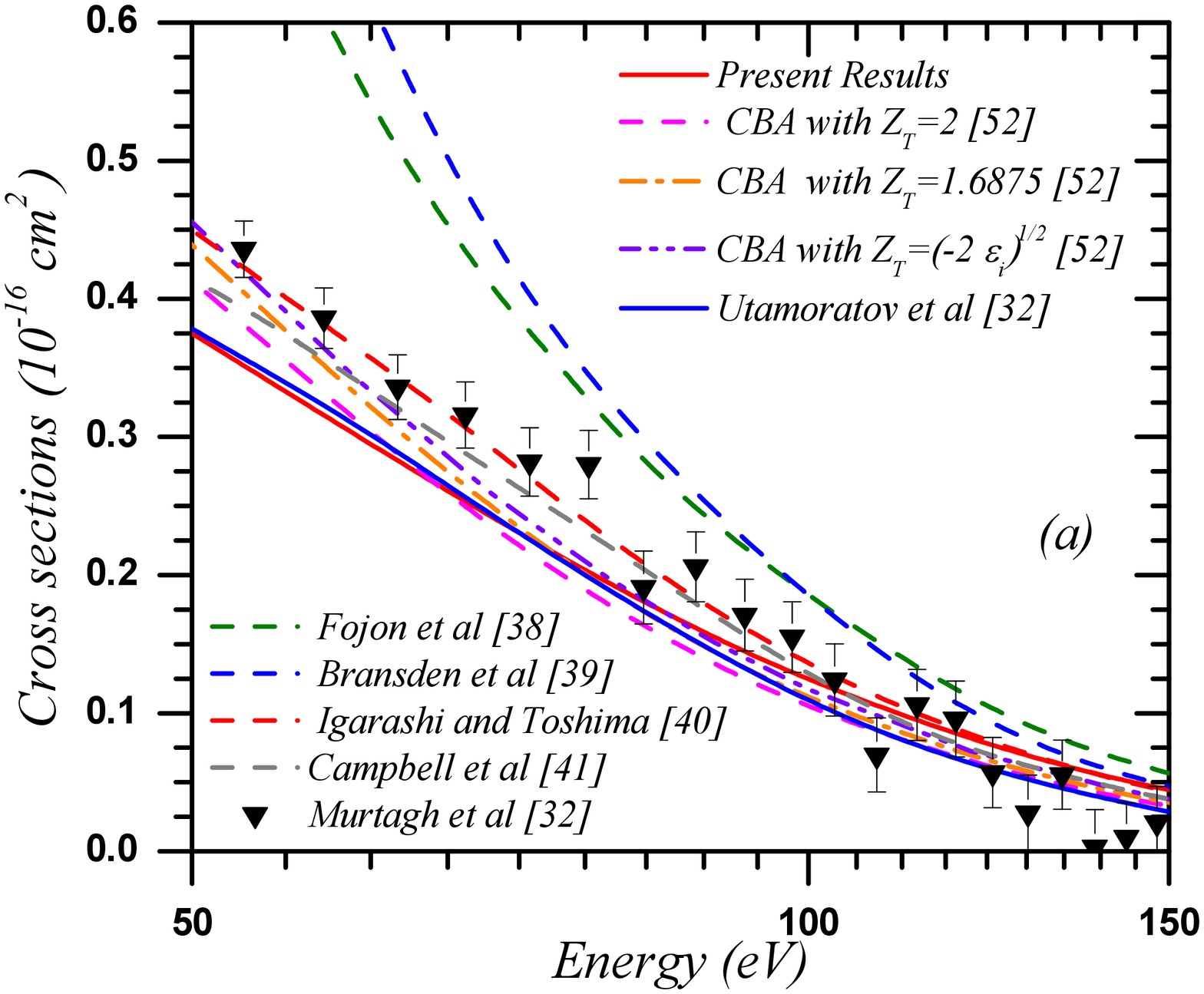}
\includegraphics[scale=0.4]{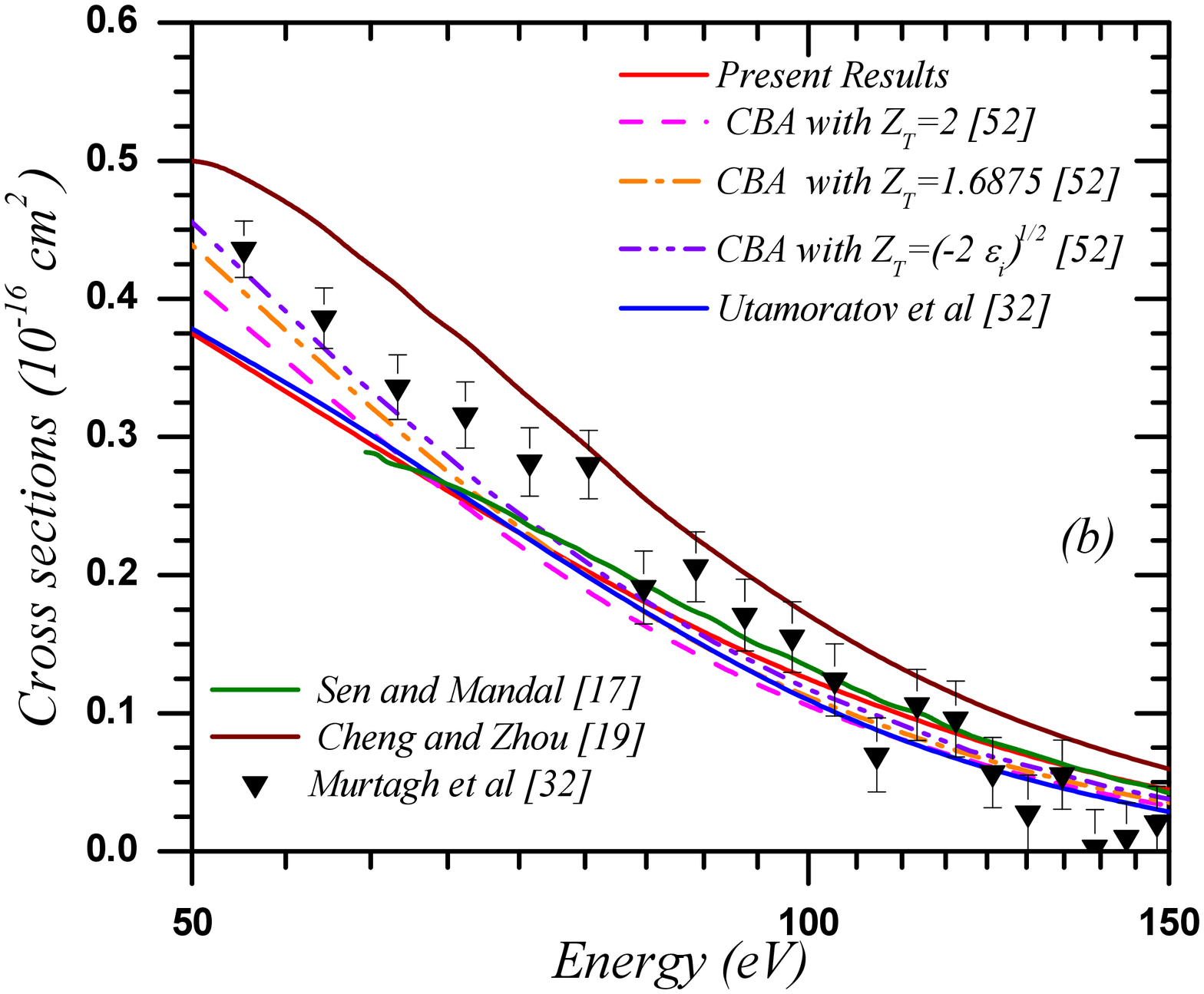}
\caption{Total cross sections for Ps formation from helium atoms
  compared the results of the listed theories and with the
experimental data by Murtagh~et~al~\cite{Murtagh2}.}\label{Fig6}
\end{figure}
Figure~\ref{Fig6} shows the total positronium-formation cross
sections in our calculation formalism and all the above-mentioned
theoretical calculations. Comparison with measurements reported by
Murtagh~et~al~\cite{Murtagh2} identifies the CDW~\cite{Igarashi},
CC~\cite{Campbell} and CBA (with $Z=2$) as the most reliable ones.
As before, our data show only fair agreement with these three
theories, at low energies, and good agreement at higher energies.
Compared with the results of Foj\'{o}n~et~al~\cite{Fojon} and of
Bransden~et~al~\cite{Bransden}, our results show substantially
superior agreement with experiment low energies. In the considered
range of impact energies, our results are compatible with those from
the more complex CCC formalism.

\section{Summary and Conclusions}
In summary, we have applied a first-order distorted wave with
correct boundary conditions to positronium formation from helium
atoms. A parameterized potential in good agreement with
Roothaan-Hartree-Fock wavefunction and the corresponding binding
energy described the screening effect of the passive electron on the
transition amplitude. Our theory is self-consistent,  since it
satisfies the correct boundary conditions and the post-prior
symmetry, and since the interaction potentials, wavefunctions,
binding energies and Coulomb phase factors in the formalism are
consistent with each other and describe the collision and electron
capture realistically.\par The chief advantage of our approach is
its simplicity. The figures in this paper comprehensively compare
the results of this approach with different sets of experimental
data and with the results of more elaborate theories. They hence set
benchmarks monitoring the accuracy of the treatment. Compared with
the results of the CBA, the angular distribution of our differential
cross sections are in qualitative agreement even at the smallest
energies: they differ quantitatively at the smallest energies, but
the agreement improves as the impact energy grows.\par Our
calculated total cross sections are in reasonable agreement with
other theories, such as the CBA, CDW, CC, CCO, and DWA, especially
at higher energies. Also the results agree well with with the
experimental data of Fornari~et~al~\cite{Fornari},
Overton~et~al~\cite{Overton}, and Murtagh~et~al~\cite{Murtagh2}, the
agreement with the latter authors being very good at incident
energies above 75~eV. As evidenced by comparison with experimental
and other theoretical data, our procedure becomes less accurate as
the impact energy is reduced, significant deviations becoming
apparent around 50~eV.\par Our relatively simple approach, based on
the first-order distorted-wave formalism  reproduces the results of
more demanding procedures, based on the second-order distorted wave
approximation. It can be applied to positronium formation in
multi-electron atoms, such a C, Ne, Na, and Ar, and also to electron
caputre from atomic shells of multi-electron atoms by bare ion
projectiles at moderate energies.

\section*{Acknowledgments} The author would like to thank Dr. Daniel
Murtagh for providing the experimental data which were measured by
his group (UCL positron group) on the different positron scattering
processes from helium atoms and Dr. Ravshan Utamoratov for
forwarding his results for Ps formation based on the two-center
convergent close-coupling formalism in tabular form.

\end{document}